\documentclass[aps,pre,floatfix,twocolumn,nofootinbib,showpacs]{revtex4} 
\usepackage{graphicx} \usepackage{amsmath} \usepackage{amssymb} 
\usepackage[sort&compress]{natbib}

\newcommand{\BEQ}{\begin{eqnarray}} 
\newcommand{\EEQ}{\end{eqnarray}} 
\newcommand{\BEA}{\begin{eqnarray}} 
\newcommand{\EEA}{\end{eqnarray}} 
 
\renewcommand{\d}{{\rm d}} 
 
\newcommand{\ep}{\varepsilon} 
 
\newcommand{\eps}{\varepsilon}

\newcommand{\Wad}{\widetilde{W}}  
\newcommand{\aad}{\widetilde{a}}

\newcommand{\half}{\frac{1}{2}}

\newcommand{\ti}{t_{\rm i}} 
\newcommand{\tf}{t_{\rm f}} 
\newcommand{\Ri}{R_{\rm i}} 
\newcommand{\Rf}{R_{\rm f}} 
\newcommand{\Hi}{H_{\rm i}} 
\newcommand{\Hf}{H_{\rm f}}
\begin{document}  
\draft 
\title 
{Minimal work principle:  proof and counterexamples} 
\author{A.E. Allahverdyan$^{1,2)}$ 
and Th.M. Nieuwenhuizen$^{1)}$} 
\address{ 
$^{1)}$ Institute for Theoretical Physics, 
Valckenierstraat 65, 1018 XE Amsterdam, The Netherlands\\ 
$^{2)}$Yerevan Physics Institute, 
Alikhanian Brothers St. 2, Yerevan 375036, Armenia}  
 
\begin{abstract}
 
The minimal work principle states that work done on a thermally
isolated equilibrium system is minimal for adiabatically slow
(reversible) realization of a given process.  This principle, one of
the formulations of the second law, is studied here for finite
(possibly large) quantum systems interacting with macroscopic sources
of work.  It is shown to be valid as long as the adiabatic energy
levels do not cross.  If level crossing does occur, counter examples
are discussed, showing that the minimal work principle can be violated
and that optimal processes are neither adiabatically slow nor
reversible. The results are corroborated by an exactly solvable model.
\end{abstract} 

 
\maketitle 

\section{Introduction}
\label{intro}

The second law of thermodynamics \cite{landau,balian,lindblad,perrot},
formulated nearly one and half century ago, continues to be under
scrutiny \cite{QL2L,AN,mahler,NA}. While its status within equilibrium
thermodynamics and statistical physics is by now well-settled
\cite{landau,balian,lindblad,perrot}, its fate in various border
situations is far from being clear
\cite{QL2L,AN,mahler,NA}. In the macroscopic realm the
second law is a set of equivalent statements concerning quantities
such as entropy, heat, work, etc.  In more general situations these
statements need not be equivalent and some, e.g. those involving
entropy, may have only a limited applicability \cite{AN,NA}.  In
contrast to entropy, the concept of work has a well-defined
operational meaning for an arbitrary system interacting with
macroscopic work sources \cite{landau,balian,lindblad}.  Moreover, the
definition of work is not confined to (nearly) equilibrium situations:
it is defined equally well for processes that start and end in
arbitrary non-equilibrium states \cite{landau,balian,lindblad}.

In this perspective the second law can be regarded as a set of several
statements concerning the work. They were first derived from
observations and have direct counterparts in everyday experience. Two
formulations of the second law are at the focus of our present
interest \cite{landau,balian,perrot,lindblad,thomson1,thomson2,
thomson3,ku,tasakif,martin,forster,fuku,narnhofer,macro}: 
\begin{itemize}

\item
Thomson's formulation: no work can be extracted
from an equilibrium system by means of a cyclic process generated by
an external work source.  

\item
The minimal work principle: when
varying the speed of a given process done on an (initially)
equilibrium system, the work is minimal for the slowest realization of
the process (more details of the principle are given below).  

\end{itemize}

These
two statements are well defined | both operationally and conceptually
| for a finite system coupled to macroscopic sources of work. It is
therefore interesting to consider whether they are valid in this
domain. This is an inquiry towards one of the most pertinent questions
in the foundations of statistical physics: whether the second law is
valid for finite systems, or, put differently, whether the
thermodynamical limit is really necessary for the validity of the
second law. Some historical developments of this question are recalled
in the Appendix.

Thomson's formulation was derived with great rigor from the 
principles of quantum mechanics \cite{thomson1,thomson2,thomson3,ku}.  It
is valid not only for a macroscopic, but also for a finite system
coupled to sources of work. This fact is not occasional: would
Thomson's formulation be violated for a finite (e.g., few-level)
system, it would immediately imply violations for 
large collections of such systems, i.e., for some macroscopic
systems.

There are also several pertinent derivations of the minimal work
principle.  They cover some important but still particular cases, such
as the linear response limit (weak coupling to the work source)
\cite{martin,forster}, quasi-slow processes \cite{fuku}, and constant
temperature processes for macroscopic systems
\cite{thomson3,narnhofer,macro,fuku}.  
So far all these derivations were
confirming the validity of the minimal work principle. 

Our present purpose is to study the minimal work principle for a
finite (possibly large) quantum system coupled to sources of work.  We
derive the principle for finite systems and then discuss its
limits. It is found that | in contrast to Thomson's formulation | the
domain of the validity of the minimal work principle is large but
definitely limited. These limits are connected with crossing of
adiabatic energy levels, and they are illustrated via counterexamples
which include an exactly solvable model. A more detailed discussion on
what, to our opinion, qualifies as a limit of the second law, is given
in the Appendix.

The phenomenon of level-crossing (conical intersections), which was
once little more than a theoretical curiosity, attracted recently much
attention in chemical physics, quantum chemistry and biophysics; see
Ref.~\cite{david} for a review. Its importance was recognized for such
essential processes as charge transfer reactions, light harvesting and
ultrafast decay of excited states.  We thus show here that the same
phenomenon of level crossing is crucial for the proper understanding
of the second law.

The paper is organized as follows. In section \ref{setup} we recall
the basic setup of the problem. Section \ref{it} discusses the minimal
work principle and its various implications. Then the principle is
derived for finite systems.  Section \ref{crossing} studies the limits
of the principle, while section \ref{crossing} discusses relations of
the principle to Thomson's formulation of the second law and to cyclic
processes.  The last section presents our conclusions.

\section{The setup.}  
\label{setup}

\subsection{Thermally isolated processes.}

Consider a quantum system S which is thermally
isolated \cite{landau,balian,perrot}: it moves according to its own
dynamics and interacts with an external macroscopic work source.
This interaction is realized via time-dependence of some parameters
$R(t)=\{R_1(t), R_2(t),...\}$ of the system's Hamiltonian
$H(t)=H\{R(t)\}$. They move along a certain trajectory $R(t)$ which at
some initial time $\ti$ starts from $\Ri=R(\ti)$, and ends at
$\Rf=R(\tf)$. The initial and final values of the Hamiltonian are
\BEA
\Hi=H\{\Ri\}\quad {\rm and}\quad \Hf=H\{\Rf\}, 
\EEA
respectively.  

The Hamiltonian $H(t)$ generates a unitary evolution: 
\BEA 
i\hbar\frac{\d}{\d t}{\rho}(t)=[\,H(t),\rho(t)\,],\qquad 
\rho(t)=U(t)\,\rho(\ti)\,U^\dagger (t), 
\label{evolution} 
\EEA 
with time-ordered 
\BEA
\label{u}
U(t)=\overleftarrow{\exp}[-\frac{i}{\hbar}\int_{\ti}^{t}\d s\, H(s)].
\EEA

The interaction of S with the source is related with flows of work
which qualifies as a high-graded (ordered or mechanical) type of energy.
The work $W$ done on S reads \cite{landau,balian,lindblad,perrot} 
\BEA 
W=\int_{\ti}^{\tf}\d t\,{\rm tr}\,[ 
\rho(t)\dot{H}(t)]={\rm tr}[\Hf\rho(\tf)]- 
{\rm tr}[\Hi\rho(\ti) ], 
\label{work} 
\EEA 
where we performed partial integration and inserted
(\ref{evolution}). This is the average energy increase of S, which,
due to energy conservation, coincides with the average energy decrease
of the source \cite{landau,balian}.  Thus there are two ways to
measure work.  One can let the work-source to interact with an ensemble of
systems S and then to measure the energy change of the
work-source. Due to the macroscopic size of the latter, its energy
practically coincides with its average \cite{landau,balian}.  Or,
alternatively, one can measure directly the initial and final average
energies of S, e.g., via measuring the Hamiltonians $\Hi$ and $\Hf$,
respectively, on the ensemble of systems S.

\subsection{Initial state.}
\label{po}

Initially the system S is assumed to be in equilibrium at temperature
$T=1/\beta\geq 0$, that is, S is described by a gibbsian density
operator:
\BEA 
\label{gibbs} 
\rho(\ti)=\frac{1}{Z_{\rm i}}\,
\exp(-\beta \Hi),
\qquad 
Z_{\rm i}={\rm tr}\,e^{-\beta \Hi}. 
\EEA 

This equilibrium state can be prepared by a weak interaction between S
and a macroscopic thermal bath at temperature $T$
\cite{landau,balian,lindblad,NA}, and then decoupling S from the bath
in order to achieve a thermally isolated process
\cite{landau,balian,lindblad,perrot}. The fact of preparing S via a
thermal bath has the following implications:
\begin{enumerate}

\item Reproducibility: The preparation process leading to $\rho(\ti)$
can in principle be repeated the needed amount of times, and thus
various effects displayed by S and described by density matrix $\rho$
can be amplified to the needed extent.  Moreover, this reproduction is
in principle not connected with energy costs: due to the weak-coupling
to the bath no (or little) energy should be spent for switching the
coupling on and off.

\item According to quantum mechanics, $\rho(\ti)$ allows to predict
probabilities for various results obtained via measurements done on S.  In
this sense $\rho(\ti)$ can be thought to describe an ensemble of
identically prepared systems S, rather than one preparation of 
a single system. Since it was
prepared via a thermal bath only |i.e., without additional
measurements employed to separate the ensemble into sub-ensembles|
no physical interpretation should be based on a choice of sub-ensembles
for $\rho(\ti)$. In particular, it will be incorrect to interpret
$\rho(\ti)$ as `` S is with some probability in a state with a
definite but unknown energy (eigenvalue of $\Hi$)''. This
interpretation is allowed for a classical Gibbs distribution, but in
the quantum situation it may lead to inconsistencies with experiment;
see, e.g., in Ref.~\cite{kok}.  One of the reasons prohibiting such an
interpretation is that separation of a mixed-state ensemble into
pure-state components (sub-ensembles) depends on the concrete measurement
done for this purpose and therefore
is not unique.

\end{enumerate}

\section{ The minimal work principle.}    
\label{it}

\subsection{Formulation of the principle.}
\label{formulation}

Let S start in the state (\ref{gibbs}), and  let
the parameters $R(t)$ move between $\Ri$ and $\Rf$ 
along a trajectory $R(t)$.  The work done on S during 
this process is $W$. 
Consider the adiabatically slow realization of this process: $R$ 
proceeds between the same values $\Ri$ and $\Rf$ and along the 
same trajectory, but now with a homogeneously vanishing speed,
thereby taking a very long time $\tf-\ti$, at the cost of 
an amount work $\Wad$.
The minimal-work principle then asserts
\cite{landau,balian,perrot}  
\BEA 
\label{2L} 
W\geq \Wad . 
\EEA 
This is a statement on optimality: if work has to be extracted from S,
$W$ is negative, and to make it as negative as possible one proceeds
with a very slow speed.  If during some operation work has to be
added ($W>0$) to S, one wishes to minimize its amount, and again
operates slowly.  

The following remarks intend to clarify the physical meaning of the
principle and to prevent its improper use.

\paragraph{}
For thermally isolated systems, adiabatically slow processes are
reversible. This is standard if S is macroscopic
\cite{landau,balian,perrot}, and below in section \ref{cyclic} it is
shown to hold for a finite S as well, where the definition of
reversibility extends unambiguously (i.e., without invoking entropy)
\cite{perrot}. Thus, the minimal work principle states that the
optimal thermally isolated processes are reversible.

\paragraph{}\label{rem1}
The formulation of the minimal work principle does not by itself give
any detailed information on the precise meaning of being
``slow''. Only its derivation from first principles can be
informative in this respect, and this is one of the reasons why such
derivations are really needed. It suffices to say at the present
moment that ``slow'' means ``slower than the characteristic times of the
system relevant for calculating work'', and certainly does not mean
``slower than characteristic times of the system S
without any interaction with external sources of work''.  Indeed, we
shall see below that in general the relevant characteristic times
become determined only once the very process $R(t)$ | with its initial
and final points $R_{\rm i}$ and $R_{\rm f}$ | is given.  The
operational definition of slow processes in the context of the minimal
work principle is straightforward: one increases the characteristic
time-scale $\tau$ of the process till the work as a function of $\tau$
saturates, i.e., it does not anymore depend on $\tau$.  It is
conceivable that for some systems there will be several type of slow
processes, that is, the work as a function of $\tau$ will display
several plateaus. If so, then each type of slow processes can be
studied for its own sake \cite{tt,at,kawasaki}.

\paragraph{}
Note that as far as the work |i.e., average energy lost by the work
source| is concerned, any processes can be considered as a part of a
thermally isolated one, provided S is defined to be the whole
system which interacts with the source of work and which together with
the latter forms a closed system 
\footnote{This is akin to the known statement that any dynamical
evolution in quantum mechanics can be viewed as a part of a unitary
evolution, provided all the environment of the system was included
into the description. Such statements are sometimes regarded as too
general and therefore useless. However, they can be very useful as
instanced by the formalism of completely positive operations for open
quantum systems \cite{lindblad}.}(see additionally in this context
Eq.~(\ref{ez}) and the remark after it). Thus, the statement of the
minimal work principle is more general than it may seem.

\paragraph{}
Finally, it may be useful to comment on the used nomenclature.
Following Ref.~\cite{landau} we call a slow thermally isolated process
adiabatic.  Now and then for sharpening the needed context we use
equivalently the term ``adiabatically slow''.  Note that this by no
means presupposes validity of the quantum-mechanical adiabatic
theorem.

\subsection{ The minimal work principle for macroscopic systems.} 
 
In macroscopic thermodynamics the minimal work principle is derived
\cite{landau,perrot} from certain axioms which ensure that, within the
domain of their applicability, this principle is equivalent to other
formulations of the second law.  Derivations in the context of
statistical thermodynamics are presented in
Refs.~\cite{narnhofer,martin,forster,fuku,macro,NA}. 

In the following discussion we will reproduce a proof of the minimal
work principle for a class of thermally isolated
processes realized on a macroscopic system.  Our purpose is to
understand why precisely the principle holds in this situation, and
thereby to motivate its investigation for finite systems.  

The derivation proceeds in two steps.
First one considers the relative entropy 
\BEA 
\label{relent}
S[\rho(\tf)||\rho_{\rm eq}(\Hf)]=
{\rm tr}[\rho(\tf)\ln\rho(\tf)-\rho(\tf)\ln\rho_{\rm eq}(\Hf)],
\EEA
between the 
final state $\rho(\tf)$ given by (\ref{evolution}) and an equilibrium 
state 
\BEA
\rho_{\rm eq}(\Hf)=\exp(-\beta \Hf)/Z_{\rm f}, \quad 
Z_{\rm f}={\rm 
tr}\,e^{-\beta \Hf}, 
\label{shun}
\EEA
which corresponds to the final Hamiltonian $\Hf$ and to the  same
temperature $T=1/\beta$ as in the initial state $\rho(\ti)$.

The relative entropy $S[\rho||\sigma]$ is known to be non-negative for
any density matrices $\rho$ and $\sigma$.  Among other useful
features, it can serve as a (non-symmetric) ``distance'' between $\rho$ and
$\sigma$, since $S[\rho||\sigma]=0$ implies
$\rho=\sigma$. Applications of relative entropy in statistical physics
and quantum information theory are reviewed in Refs.~\cite{balian,vedral}.

As follows from the unitarity of the evolution
operator (\ref{evolution}), 
\BEA
S_{\rm f}=-{\rm tr}[\rho(\tf)\ln\rho(\tf)]= -{\rm
tr}[\rho(\ti)\ln\rho(\ti)]=S_{\rm i}.
\label{dard}
\EEA
This is the well-known conservation of Von Neumann entropy during
a thermally isolated process.

The definition of relative entropy (\ref{relent}), combined with 
Eqs.~(\ref{dard}, \ref{shun}, \ref{gibbs}) and with the definition of
work (\ref{work}) brings:
\BEA
S[\rho(\tf)||\rho_{\rm eq}(\Hf)]=
{\rm tr}[\rho(\ti)\ln\rho(\ti)-\rho(\tf)\ln\rho_{\rm eq}(\Hf)]\nonumber\\
=\beta W+\ln Z_{\rm f}-\ln Z_{\rm i}.~~~
\EEA

Using the
non-negativity of the relative entropy (\ref{relent}) one gets
\BEA 
\label{dubonos}
W=&& F(\Hf)-F(\Hi)+T\,S[\rho(\tf)||\rho_{\rm eq}(\Hf)]
\\
\geq&& F(\Hf)-F(\Hi)
\label{workfree-energy} 
\EEA 
where $F(H)$ is the free energies corresponding 
to Hamiltonian $H$ and temperature $T$:
\BEA
F(H)\equiv
-T\,\ln {\rm tr}\,e^{-\beta H}.
\EEA
This inequality is well-known \cite{landau,balian} 
(though is derived and formulated less explicitly)
and sometimes is viewed as a proof of the minimal
work principle \footnote{Recently inequality (\ref{workfree-energy}) 
was generalized
within so called work-fluctuations theorems \cite{flu,ku}.
These theorems account for fluctuations which appear when
measuring the work via the system's energy.
}. 
However, we still have to show that the work $\Wad$
in the slow limit coincides with the difference
of free energies given by the right hand side
(RHS) of Eq.~(\ref{workfree-energy}).
Recall that the latter quantity was so far defined only formally.
 
Eq.~(\ref{workfree-energy}) gets the needed physical meaning 
when one {\it assumes} that for a macroscopic
system S, the final state $\tilde{\rho}(\tf)$ |reached from
$\rho(\ti)=\exp(-\beta \Hi)/Z_{\rm i}$ by the adiabatically slow
process| can be approximated by $\rho_{\rm eq}(\Hf)$ defined in 
Eq.~(\ref{shun}):
\BEA
\label{shuka}
S[\tilde{\rho}(\tf)||\rho_{\rm eq}(\Hf)]\approx 0,
\EEA
where $\approx$ means that the equality is supposed to recover in the 
macroscopic (thermodynamic) limit for S.
This then leads from Eq.~(\ref{dubonos})
to the needed relation
\BEA
\label{ddum}
\Wad \approx F(\Hf)-F(\Hi).
\EEA
The statement (\ref{2L}) of the minimal work principle
then follows from Eqs.~(\ref{dard}, \ref{ddum}).

In the second step of the derivation one should prove 
Eq.~(\ref{ddum}).  There are several classes of macroscopic systems
for which one can show that the free energy difference $F(\Hf)-F(\Hi)$
in (\ref{workfree-energy}) indeed coincides with the adiabatic work~
\cite{narnhofer,macro,NA}.  We recall below one of them.

Assume that S consists of two parts: macroscopic thermal bath B and a
subsystem (particle) P coupled to it:
\BEA
\label{kov}
H(t)=H_{\rm P}\{R(t)\}+H_{\rm B}+g\,H_{\rm I},
\EEA
where $H_{\rm P}(t)$ and $H_{\rm B}$ stand for the Hamiltonians of P
and B, respectively, and where $H_{\rm I}$ is the interaction
Hamiltonian with $g$ being the corresponding coupling
constant. The source of work interacts with the particle only, thus
only $H_{\rm P}\{R(t)\}$ is time-dependent. 

The following conditions are usually considered to be sufficient for the 
validity of Eq.~(\ref{ddum}); see \cite{landau,balian}. 
Along the lines of Ref.~\cite{NA}, we present them in a slightly more 
formalized and particular way,
and we recall that they were checked in models \cite{NA,macro}.

\begin{itemize}

\item
The thermal bath B is composed of a macroscopic number of harmonic
oscillators with a proper (e.g. ohmic) spectrum of their interaction
with P. 

\item
The evolution generated by $H(t)$ in (\ref{kov})
starts from an equilibrium state (\ref{gibbs}) for the
total system S=B+P.

\item 
The characteristic time of the external process $R(t)$ is assumed to
be much larger than the relevant times of the particle P. These times
include those generated by the Hamiltonian $H_{\rm P}\{R(t)\}$, as
well as the relaxation times of P induced by the bath. The latter
times are controlled by the interaction Hamiltonian $H_{\rm I}$ and
they become very long for $g\to 0$. Note that since the thermal bath B
is assumed to be macroscopic (dense spectrum), there are
characteristic times of B (so called Heisenberg times)
which are proportional to inverse
level-spacing and are thus very large. They, however, do not enter
into the definition of adiabatically slow processes.  All the relevant
characteristic times are finite in the thermodynamical limit for the
bath, at least for the type of models considered in \cite{NA}.

\end{itemize}

The above three conditions are sufficient for
the density operator of P to be given as \cite{NA}
\BEA
\label{barbudos}
\rho_{\rm P}(t)=\frac{1}{Z(t)} {\rm tr}_{\rm B}\,{e^{-\beta
H(t)}},\qquad Z(t)={{\rm tr}\, e^{-\beta H(t)}}, \EEA where ${\rm
tr}_{\rm B(P)}$ means trace over the bath (particle) degrees of freedom.  
The physical meaning of (\ref{barbudos}) is obvious: in the slow limit 
P is in the local-equilibrium (or local-stationary) state
\footnote{For the actual calculation in Eq.~(\ref{barbudos}), one may need to
keep the bath large but finite, to carry out the trace ${\rm tr}_{\rm
B}$, and only then to go to the macroscopic limit for the bath. Then
the density matrix $\rho_{\rm P}(t)$ is finite and well-behaved
although the quantities like $Z(t)={{\rm tr}\, e^{-\beta H(t)}}$ may
not be well-defined in the macroscopic limit for the bath. This is a
standard procedure in statistical physics and it is legitimate for the
present situation, since the thermodynamical limit for the bath
commutes with the limit of slow processes, as we already recalled
above. The validity of such commutation was also seen in
Ref.~\cite{fuku}.}.

Note that the state of P at time $t$ need not at all coincide 
with the local Gibbsian $e^{-\beta H_{\rm P}(t)}/Z_{\rm P}(t)$, 
since no weak-coupling
assumption on the system-bath interaction was made.

Eqs.~(\ref{work}, \ref{kov}, \ref{barbudos}) suffice to derive
relation (\ref{ddum}) between the work $\Wad$ done for the
adiabatically slow process and the difference $F(\Hf)-F(\Hi)$ of free
energies. Indeed, note from Eq.~(\ref{kov}) that $\partial _tH(t)
=\partial _tH_{\rm P}(t)$ and (\ref{work}) can be written as
\BEA
\label{ez}
W=\int_{\ti}^{\tf}\d t\,{\rm tr}_{\rm P}\,[ 
\rho_{\rm P}(t)\,\,\partial _t
H_{\rm P}(t)].
\EEA
This relation shows that the work defined globally via the energy
difference of the overall system of the particle and the bath can also
be calculated via integration of the expression in the RHS of
Eq.~(\ref{ez}), which contains only quantities referring to the
particle P.

Now proceed with application of Eq.~(\ref{barbudos}):
\BEA
W=&&
\int_{\ti}^{\tf}\frac{\d t}{Z(t)}\,
{\rm tr}_{\rm P}\,{\rm tr}_{\rm B}\,
\left[ 
\,e^{-\beta H(t)}\,\,\partial _tH_{\rm P}(t)\,\right]\nonumber\\
=&&
\int_{\ti}^{\tf}\frac{\d t}{Z(t)}\,{\rm tr}\,\left[ 
\,e^{-\beta H(t)}\,\,\partial _tH(t)\,\right]
\nonumber\\
=&&-T
\int_{\ti}^{\tf}\d t\,\frac{\d }{\d t}\,\ln Z(t)=F(\Hf)-F(\Hi).
\EEA

Thus the minimal work principle (\ref{2L}) is proved for this class of 
processes realized on macroscopic systems. One may note once again that 
although the quantities $F(\Hf)$ and $F(\Hi)$ are very large
in the macroscopic limit for the bath, their difference is finite
and is order of the particle's energy. This is also seen from (\ref{ez})
which contains only quantities referring to the particle.
 
\section{ Finite systems.} 

Let us now turn to a finite $N$-level quantum system S.  
\footnote{Note that an attempt was made recently to study the minimal
work principle for finite systems \cite{tasakif}. However, the result
of  this study 
is to our opinion incomplete, since the author of Ref.~\cite{tasakif}
obtained that the principle has the same range of validity as Thomson's 
formulation of the second law. This came from incorrect treatment 
of adiabatically slow processes. It is however to be mentioned that 
the author of Ref.~\cite{tasakif} has stressed the preliminary character
of his results, and that his results concerning cyclic processes are
correct.}
The first thing to do is to apply here the reasoning developed above
for macroscopic systems. In fact, Thomson's formulation of the second
law, $W\geq 0$, is valid for finite systems for precisely the same
reasons as it is valid for macroscopic ones. Indeed, since this
formulation refers to cyclic processes, one puts $H_{\rm f}=H_{\rm i}$
in Eqs.(\ref{relent}, \ref{workfree-energy}) | this can be done
without altering the validity of (\ref{workfree-energy}), since
$\rho_{\rm eq}$ is in general an auxiliary density matrix | and gets
$W\geq 0$ for cyclic processes, which is the statement of Thomson's
formulation.

However, when trying to apply the above reasoning for finite systems, one
immediately sees that there are no reasons why the work in the slow limit
should be equal to the difference in free energies.  Neither are there
reasons to expect validity of Eq.~(\ref{shuka}), that is, to expect
that in general the final state $\tilde{\rho}(\tf)$ reached
from $\rho(\ti)=\exp(-\beta \Hi)/Z_{\rm i}$ during the adiabatically
slow process, should be equal to $\rho_{\rm eq}(\Hf)$ defined by
(\ref{shun}). The reason for this can be seen by noting that due to
the unitarity of the evolution operator $U$ in (\ref{u}), the spectrum
of $\tilde{\rho}(\tf)$ coincides with that of the initial density
matrix $\rho(\ti)$, and in general does not have the Gibbsian shape of
the spectrum of $\rho_{\rm eq}(\Hf)$.

The facts that $\tilde{\rho}(\tf)\not= \rho_{\rm eq}(\Hf)$ and $\Wad
\not =F(\Hf)-F(\Hi)$ for finite and certain macroscopic systems are
known and were studied especially in the linear response regime
 \cite{klein,rosen,mountain,wilcox,mashka,japan}. Here are some
results obtained within that activity: {\it i)} A class of finite
systems was determined for which the adiabatic work coincides with the
difference in free energies \cite{klein,rosen}. {\it ii)} Conditions
were determined for macroscopic systems, under which the adiabatic
work converges to the free energy difference in the thermodynamical
limit \cite{rosen,mountain,mashka}. Concrete estimates for the rate
of convergence were given. {\it iii)} It was found that for
certain non-trivial macroscopic systems the above convergence can be
absent even in the thermodynamical limit \cite{rosen,wilcox}.  It was
argued that quantum systems are more vulnerable in this respect
than classical ones \cite{rosen}. {\it iv)} The fact that
$\tilde{\rho}(\tf)\not= \rho_{\rm eq}(\Hf)$ was recently applied for
studying certain processes which are reversible in the macroscopic
limit, but become irreversible for finite system  \cite{japan}.

Thus we cannot rely on macroscopic analogies and
we need an independent  derivation of the minimal work principle
(\ref{2L}) for finite systems. 
Some ideas of the following derivation were adopted 
from Ref.~\cite{thomson2}. Let the spectral resolution 
of $H(t)$ and $\rho(\ti)$ be 
\footnote{The eigenstates of $\rho(\ti)$ are used below as a calculational
tool. It is by no means implied that ``what really happens'' is that the
system S is |with some probability|
in one of those states. See also
the second remark in section \ref{po}.}
\BEA 
H(t)=\sum_{k=1}^N\ep_k(t)|k,t\rangle\langle k,t|,\quad 
\langle k,t|n,t\rangle =\delta_{kn},\\ 
\label{khorezm} 
\rho(\ti)=\sum_{k=1}^N p_k|k,\ti\rangle\langle k,\ti|,\quad 
p_k=\frac{e^{-\beta \ep_k(\ti)}}{\sum_n e^{-\beta \ep_n(\ti)}}. 
\EEA 
At $t=\ti$ we order the spectrum as 
\BEA 
\label{jalaledin}
&&\ep_1(\ti)\leq ...\leq \ep_N(\ti)\,
\Longrightarrow\\
&&p_1\geq...\geq p_N. 
\label{jalaledin'} 
\EEA 

For any $t$ in the interval
$\ti\leq t\leq \tf$ we expand over the complete set $|n,t\rangle$: 
\BEA 
\label{chingiz} 
U(t)|k,\ti\rangle=\sum_{n=1}^N a_{kn}(t) \, 
e^{-\frac{i}{\hbar}\int_{\ti}^{t}\d t'\,\ep_n(t')}\,  
|n,t\rangle, 
\EEA 
where 
\BEA
a_{kn}(t)\equiv a_{kn}(t;\ti) = 
\langle n,t| \,U(t)|k,\ti\rangle\,
e^{\frac{i}{\hbar}\int_{\ti}^{t}\d t'\,\ep_n(t')},\nonumber\\
\label{marz}
\EEA
are the expansion coefficients, and where $|a_{kn}(\tf)|^2$ is
the probability to measure energy $\ep_n(\tf)$ at the final time $\tf$,
provided the initial state was $|k,\ti\rangle\langle k,\ti|$.  

Now we use Eqs.~(\ref{khorezm}, \ref{chingiz}) to obtain for the work
(\ref{work}): 
\BEA 
\label{suomi0} 
W=\sum_{k,n=1}^N 
|a_{kn}(\tf)|^2
\,p_k\,\ep_n(\tf)-\sum_{k=1}^N p_k\,\ep_k(\ti). 
\EEA 
A similar formula can be derived to express the adiabatic work
$\Wad$ in terms of coefficients  $\aad_{kn}(\tf)$:
\BEA 
\label{suomi0'} 
\Wad=\sum_{k,n=1}^N 
|\,\aad_{kn}(\tf)|^2\,p_k\,\ep_n(\tf)-\sum_{k=1}^N p_k\,\ep_k(\ti). 
\EEA 

From the definition (\ref{marz}),
\BEA
|a_{kn}(\tf)|^2=|\langle n,\tf|U|k,\ti\rangle|^2,
\EEA 
it follows that \footnote{
As possible physical interpretation of feature
(\ref{bek}), note that for the uniform distribution of the
initial states, $p(k,\ti)=\frac{1}{N}$, the prediction
probability $p(n;\tf\,|\, k,\ti)=|a_{kn}(\tf)|^2$
is equal to the retrodiction probability
$p(k,\ti \,|\,n;\tf)=\frac{p(k,\ti)\,p(n;\tf\,|\, k,\ti)}
{\sum_{k=1}^Np(k,\ti)\,p(n;\tf\,|\, k,\ti)}=
\frac{\frac{1}{N}\,p(n;\tf\,|\, k,\ti)}
{\sum_{k=1}^N\frac{1}{N}\,p(n;\tf\,|\, k,\ti)}
=p(n;\tf\,|\, k,\ti)
$.
}
\BEA 
\label{bek} 
\sum_{k=1}^{N}|a_{kn}(\tf)|^2=\sum_{n=1}^{N}|a_{kn}(\tf)|^2=1. 
\EEA 

Employing the identity (summation by parts): 
\BEA
\sum_{n=1}^N \ep_n x_n=\ep_N\sum_{n=1}^Nx_n- 
\sum_{m=1}^{N-1}[\ep_{m+1}-\ep_{m}]\sum_{n=1}^m x_n, 
\EEA
with $x_n=\sum_{k=1}^N |a_{kn}(\tf)|^2\,p_k$ and
$x_n=\sum_{k=1}^N |\aad_{kn}(\tf)|^2\,p_k$,
we obtain
from Eqs.~(\ref{suomi0}, \ref{suomi0'}) and using Eqs.~(\ref{bek}) a 
general formula for the difference between the non-adiabatic 
and adiabatic work:
\BEA 
\label{suomi1} 
W-\Wad=&&\sum_{m=1}^{N-1}[\ep_{m+1}(\tf)-\ep_{m}(\tf)]\,\Theta_m,\\ 
\Theta_m\equiv&&\sum_{n=1}^{m}\sum_{k=1}^N p_k 
(\,|\aad_{kn}(\tf)|^2-|a_{kn}(\tf)|^2). 
\label{suomi2} 
\EEA 

Let us now assume that the ordering (\ref{jalaledin}) is kept at the 
final time $t=\tf$: 
\BEA 
\label{20'} 
\ep_1(\tf)\leq ...\leq \ep_N(\tf). 
\EEA 
If different energy levels did not cross each other
|i.e., different ones do not become equal and
equal ones do not become different | Eq. (\ref{20'})
is implied by the initial ordering (\ref{jalaledin}).  

The behavior of energy levels with respect to level-crossing is governed
by the non-crossing rule, which was numerously discussed in literature
and derived in a rather general situation \cite{nw,landauquant,mead1,mead11}.
We shall need it in the following particular formulation:
\begin{itemize}
\item 
If $H\{R\}$ is real and only one of its parameters is varied with
time, (\ref{20'}) is satisfied for any discrete-level quantum system:
level-crossing, even if it happens in model-dependent calculations or
due to approximate symmetry, does not survive arbitrary small
perturbation where it is substituted by avoided crossing (for a more
general $H\{R\}$ the conditions prohibiting level-crossing are more
restrictive; see \cite{mead1}).  To get a stable point of
level-crossing, one needs at least two independently varying
parameters for a real $H\{R\}$.
\end{itemize}

The rule is known since the early days of quantum mechanics \cite{nw}
and is presented in textbooks; see, e.g., Ref.~\cite{landauquant}.  It
is an important tool in atomic and molecular spectroscopy, and
development of these fields led people to reconsider its derivation.
In particular, Ref.~\cite{hutton} correctly criticizes the standard
proof of the no-crossing rule for being not precise and not general
enough. In response to this and several related criticisms, (at least)
two complete and general derivations appeared which settled the issue
 \cite{mead1,mead11}.

No level-crossings and natural conditions of smoothness of $H(t)$
are {\it sufficient} for the standard quantum adiabatic theorem 
\cite{standardad} to ensure
\BEA 
\label{abu} 
\aad_{kn}(\tf)=\delta_{kn}. 
\EEA 
This is the known statement on the absence of transitions for slow
variations \footnote{\label{times}
This standard statement of the adiabatic theorem
was elaborated in literature several times so as to provide information
on the internal characteristic times ${\cal T}$ \cite{pechu}. To be slow
then will mean $\tau\gg {\cal T}$. Dealing with a discrete spectrum
and assuming, as we did above, that there are no level-crossings and
that adiabatic energy levels are smooth functions of time, ${\cal T}$
can roughly be estimated via the inverse of the minimal spacing between
the involved energy levels \cite{pechu}. In this context, the levels
which provide a small spacing are said to define avoided crossing.
One should however keep in mind that this estimation cannot be
extrapolated to situations with level crossing, where the minimal
level-spacing is zero. In those cases the whole situation changes and
the characteristic internal times ${\cal T}$ can well be finite; see
below and in Refs.~\cite{hage,avron} (compare also with our discussion
in section \ref{formulation}).  The same is the case for some situations in
a continuous spectrum, where again the minimal level-spacing is zero, but the
relevant characteristic times can be finite; see, e.g., 
\cite{simon,elgart}. In short: the inverse of
the minimal level-spacing defines the relevant characteristic time only
for some particular cases, not in general.}. 

Combined with Eqs.~(\ref{jalaledin'}, \ref{abu}), the definition
(\ref{suomi2}) brings  
\BEA 
\label{khan} 
\Theta_m=&&\sum_{k=1}^{m}p_k\left[1-\sum_{n=1}^m |a_{kn}(\tf)|^2\right]
\nonumber\\
-&&\sum_{n=1}^{m}\sum_{k=m+1}^N p_k|a_{kn}(\tf)|^2\quad\nonumber\\ 
\geq&& -p_m\left[\sum_{k=1}^{m}\sum_{n=1}^m |a_{kn}(\tf)|^2+
\sum_{n=1}^{m}\sum_{k=m+1}^N |a_{kn}(\tf)|^2\right]\nonumber\\
&&+p_m m=p_m(m-m)
=0.
\EEA 
Once $\Theta_m$ are non-negative, the statement $W\geq \Wad$
of the minimal work principle follows from
Eqs.~(\ref{suomi0}, \ref{suomi1}, \ref{20'}). 

Recall once again the basic ingredients of the proof: 
\begin{itemize}

\item The condition (\ref{bek}) which came from the unitarity
of the evolution during a thermally isolated process. 

\item The no-crossing assumption which lead to Eq.~(\ref{20'})
and simultaneously | via the standard adiabatic theorem |
to Eq.~(\ref{abu}). 

\item We did not assumed that the adiabatic work is either related or
equal to the difference in free energies.

\item Only two features of the initial state $\rho(\ti)$ were 
used: $[H_{\rm i}, \rho(\ti)]=0$ determined the specific form 
(\ref{suomi0}, \ref{suomi0'}) of the work, while Eq.~(\ref{jalaledin'})
was used in proving $\Theta_m\geq 0$ in Eq.~(\ref{khan}). 

\end{itemize}

As an immediate application of the obtained results note that 
if only one parameter is varied, we are ensured of the absence of level
crossings and the minimal work principle is valid. Among many examples
of this situation there is the case of a gas | which may consist of
any number of particles | in a rectangular container interacting with
one of its walls moving in time (the standard setup for a
one-parameter thermally isolated process).
 
\section{Level crossing.} 
\label{crossing}

The above non-crossing condition raises 
the question: Is the minimal work principle
also valid if the adiabatic energy levels cross?  
Before addressing this question in detail, let us mention some
popular misconceptions which surround the level-crossing problem: 

| The no-crossing rule is said to exclude all crossings. This is
incorrect as the exclusion concerns situations where, in particular,
only one independent parameter of a real Hamiltonian $H\{R\}$ is
varied \cite{mead1}. Two parameters can produce robust level-crossing
for such Hamiltonians.

| One accepts that level-crossing can happen, but believes it to be
very rare and thus irrelevant for any sensible physical
situation. This opinion is invalidated by the whole chapters of
chemical physics \cite{david,chem}: level crossing is not only
observed in many--(at least two)--atom molecules, where internuclear
distances play the role of classical time-dependent parameters, but is
necessary for the proper description of some known chemical reactions,
as well as for predicting new ones. Over the years several methods
were developed for identifying and locating the points of
level-crossing. 
\footnote{This is an important issue, since some numerical or approximate
analytical methods may easily miss points of level crossing or may
produce spurious ones.}  In particular, the method based on the
geometrical phase (``Berry phase'') allows to deduce the existence of level-crossing
from the behavior of the system at points remote from the
crossing; see Refs.~\cite{david} for more information.

Note as well that level-crossing is a more frequent phenomenon
than avoided crossing~ \cite{rare}, if the number of
independently varying parameters is larger than two.

| It is believed that once levels can cross, $\Delta\ep\to 0$, the
very point of defining slow processes disappears as the internal
characteristic time $\hbar/\Delta\ep$ of S is infinite. This view
misidentifies the proper internal time as seen below (see also the
discussion in section \ref{formulation} and Footnote \ref{times}).  
The absence of
level-crossings is indeed a sufficient condition for the validity of
Eq.~(\ref{abu}) (no transitions for slow changes). It is however by no
means necessary \cite{bornfock,hage,avron}.

| It is sometimes believed that crossing is automatically followed by
a population inversion. We shall find no support for that.

\subsection{A two-level example within adiabatic perturbation theory.}

As a first example we consider a spin-$1/2$ particle  
with Hamiltonian 
\BEA 
\label{sos} 
H(s)&&=h_1(s)\sigma_1-h_3(s)\sigma_3,\\
&&=\left(\begin{array}{rr} 
-h_3(s)&h_1(s) \\ 
h_1(s)&h_3(s) 
\end{array}\right),
\qquad 
s=t/\tau, 
\label{soso}
\EEA 
where $\sigma_1$, $\sigma_3$ and $\sigma_2=i\sigma_1\sigma_3$ are
Pauli matrices, and where $s$ is the reduced time with $\tau$ being
the characteristic time-scale.  The magnetic fields $h_1$ and $h_3$
smoothly vary in time. 

Let us now make the following assumptions {\it i)} for $s\to s_{\rm
i}<0$ and for $s\to s_{\rm f}>0$, $h_1(s)$ and $h_3(s)$ go to constant
values sufficiently fast. {\it ii)} At $s=t=0$ both $h_1(s)$ and
$h_3(s)$ turn to zero. This conditions indicates a level crossing,
since crossed eigenvalues of the traceless $2\times 2$ matrix $H(s)$
means that this matrix is equal to zero at the point of the
crossing. Besides these two basic conditions, we shall assume few
auxiliary ones, whose purpose is to make the discussion below more
transparent.  {\it iii)} $h_1(s)$ and $h_3(s)$ are non-zero for all
$s$, $s_{\rm i}\leq s\leq s_{\rm f}$, except $s=0$.  {\it iv)} Due to
the latter condition $\frac{h_3(s)}{h_1(s)}$ is finite for the
involved $s$, except possibly $s\to 0$. If it infinite in this limit,
then obviously its inverse $\frac{h_1(s)}{h_3(s)}$ is finite, now for
all $s$, $s_{\rm i}\leq s\leq s_{\rm f}$, and goes to zero for $s\to
0$. Since one of these ratios has to be finite, we shall assume that
this finite ratio is $\frac{h_1(s)}{h_3(s)}$.  {\it v)} $h_3(s)>0$ for
$s<0$ and $h_3(s)<0$ for $s>0$.

Here is a concrete example realizing the last two assumptions. For
$s\to 0$: 
\BEA
h_1(s)\simeq
\alpha_1 s^2, 
\qquad
h_3(s)\simeq -\alpha_3 s, 
\label{durman}
\EEA
where $\alpha_1$ and $\alpha_3$ are positive constants. 

With the above conditions on $h_1(s)$ and $h_3(s)$, one can propose a
useful parametrization of the Hamiltonian (\ref{soso}). Recalling that
$h_3(s)$ changes its sign at $s=0$, Eq.~(\ref{soso}) is written as
\BEA 
\label{dalal}
&&H(s) = 
{\rm sg}(s)\sqrt{h^2_3(s)+h^2_1(s)}\,
\left(\begin{array}{rr} 
\cos\theta(s)&\sin\theta(s) \\ 
\sin\theta(s)&-\cos\theta(s) 
\end{array}\right),\nonumber\\
&&
\EEA
where
\BEA
&&\sin\theta(s)\equiv{\rm sg}(s)\frac{h_1(s)}{\sqrt{h^2_3(s)+h^2_1(s)}},\\
&&\cos\theta(s)\equiv-{\rm sg}(s)\frac{h_3(s)}{\sqrt{h^2_3(s)+h^2_1(s)}}.
\EEA 
Here $\sqrt{\dots}$ is defined to be always positive,
${\rm sg}(s)$ is the sign function and
\BEA
\theta(s)=-\arctan\left[\frac{h_1(s)}{h_3(s)}\right],
\label{mushik1}
\EEA
is a parameter in the interval $-\pi/2<\theta(s)<\pi/2$. Note that the
presence of ${\rm sg}(s)$ in the above expressions is necessary for
having a smooth parametrization. Otherwise, e.g.,
${h_3(s)}/{\sqrt{h^2_3(s)+h^2_1(s)}}$ is discontinuous at the crossing
point $s=0$.

The eigenvalues and eigenvectors of $H(s)$,
\BEA
H(s)|k,s\rangle=\ep_k(s)|k,s\rangle,\quad k=1,2,
\EEA 
are now read off by inspecting Eq.~(\ref{dalal}):
\BEA 
\label{mushik2}
\ep_1(s)={\rm sg}(s)\sqrt{h^2_3(s)+h^2_1(s)},\quad 
\ep_2(s)=-\ep_1(s), 
\EEA 
\BEA 
|1,s\rangle= 
\left(\begin{array}{r} 
\cos\frac{1}{2}{\theta(s)} \\ 
\\ 
\sin\frac{1}{2}{\theta(s)}  
\end{array}\right),\quad 
|2,s\rangle= 
\left(\begin{array}{r} 
-\sin\frac{1}{2}{\theta(s)} \\ 
\\ 
\cos\frac{1}{2}{\theta(s)}  
\end{array}\right). 
\label{sosend} 
\EEA 
It is seen that the eigenvalues $\eps_1(s)$ and $\eps_2(s)$
of $H(s)$  (adiabatic energy levels)
cross at $s=\theta(s)=0$.
Note from the above conditions on $h_1(s)$ and $h_2(s)$ and from
Eqs.~(\ref{mushik1}, \ref{mushik2}) that both the eigenvalues and the
eigenvectors of $H(s)$ are smooth functions of $s$.  This fact is
important for subsequent calculations, but first of all it is
necessary for the very definition of level-crossing.

Now Eq.~(\ref{jalaledin}) for the ordering of initial energy levels is
valid, but Eq.~(\ref{20'}) for the same ordering of final energy
levels is not valid due to the level-crossing at $s=0$.
Eqs.~(\ref{suomi0}--\ref{suomi2}) imply:
\BEA \label{WminWadcross}
W-\Wad=-2\sqrt{h^2_1(s_{\rm f})+h^2_3(s_{\rm f})}\,\Theta_1, 
\quad \tau\,s_{\rm f}=\tf, 
\EEA 
where $\Theta_1$ is defined by (\ref{suomi2}).

Our strategy is now to confirm relation (\ref{abu}) in the slow limit  
$\tau\to \infty$ and thus to confirm that $\Theta_1>0$,
implying that the minimal work principle is indeed violated.  

To this end we apply the standard quantum mechanical
adiabatic perturbation 
theory \cite{standardad}. Substituting Eq.~(\ref{chingiz}) into  
Eq.~(\ref{evolution}) one has: 
\BEA 
&&\dot{a}_{kn}=-\sum_{m=1}^N a_{km}(t)\,
e^{\frac{i}{\hbar}\int_{\ti}^t\d t'[\ep_{n}(t')-\ep_{m}(t')]
}\, 
\langle n,s|\partial_t|m,s\rangle.\nonumber\\
&&
\label{shaman} 
\EEA 

As $|1,s\rangle$ and $|2,s\rangle$ in Eq.~(\ref{sosend}) are real, 
the normalization
$\langle n,s|n,s\rangle=1$ implies $\langle n,s|\partial_t|n,s\rangle=0$. 
Since $\langle n,s|\partial_t|m,s\rangle 
=\frac{1}{\tau}\langle n,s|\partial_s|m,s\rangle$ the RHS of (\ref{shaman}) 
contains a small parameter ${1}/{\tau}$. 
It is therefore more convenient to introduce new variables: 
$a_{kn}(t)=\delta_{kn}+b_{kn}(t)$, $b_{kn}(\ti)=0$. 
To leading order in $1/\tau$,
$b_{kn}$ can be neglected in the RHS of (\ref{shaman}), 
and after changing variables as 
$s\tau=t$, $s'\tau=t'$, one gets for
$a_{k\not =n}(\tf)=b_{k\not =n}(\tf)$:
\BEA 
|a_{k\not =n}(\tf)|^2
=\left|\int_{s_{\rm i}}^{s_{\rm f}}\d s\,e^{ 
\frac{i\tau}{\hbar}\int^s_{s_{\rm i}}\d u[\ep_k(u)-\ep_n(u)]} 
\langle n|\partial_s|k\rangle\right|^2,~ 
\label{w3} 
\EEA 
while, due to normalization, 
\BEA
|a_{kk}(\tf)|^2=1-\sum_{n\not =k}|a_{kn}(\tf)|^2.
\EEA 

For our model described by Eqs.~(\ref{sos}--\ref{sosend}), 
the quantity
\BEA
\label{kra}
\int_{s_{\rm i}}^s\d u[\ep_1(u)-\ep_2(u)]\\
=2\int_{s_{\rm i}}^s\d u\,\ep_1(u)
\EEA
has only one extremal point, at $s=0$.  
We also have from (
\ref{sosend}) 
\BEA 
\langle 2,s|\partial_s|1,s\rangle= 
\frac{\theta'(s)}{2}=\half\,\frac{h_1(s)h'_3(s)-h_3(s)h'_1(s)}
{h^2_3(s)+h^2_1(s)},
\EEA
where $\theta'(s)\equiv{\d \theta}/{\d s}$. 

For large $\tau$ the integral in Eq.~(\ref{w3}) can be calculated with use
of the saddle-point method:
\BEA 
|a_{12}(\tf)|^2 
=\frac{\pi\hbar}{\tau}\left.
\left[\,
\frac{\langle 2,s|\partial_s|1,s\rangle^2
\,\sqrt{h^2_1(s)+h^2_3(s)}}
{h_1(s)h'_1(s)+h_3(s)h'_3(s)}\,\right]\right|_{s=0}.\nonumber\\
\label{w7} 
\EEA 
Substituting Eq.~(\ref{durman}) into
Eq.~(\ref{w7}), one gets
\BEA
\label{w77} 
|a_{12}(\tf)|^2 
=\frac{\pi\hbar\alpha_1^2}{4\tau\alpha_3^3}.
\EEA

Eqs.~(\ref{w3}, \ref{w7}, \ref{w77}) extend the
statement of the adiabatic theorem (\ref{abu}) for
the level-crossing situation. More general versions of similar adiabatic
theorems can be found in Refs.~\cite{hage,avron}.  Inserting
\BEA
\Theta_1=(p_1-p_2)|a_{12}(\tf)|^2>0 
\EEA
in Eq.~(\ref{WminWadcross})
confirms the violation of the minimal work principle. 
Eq.~(\ref{w77}) also shows that for the considered process the role of
the proper internal characteristic time is played by $\hbar
\alpha_1^2/\alpha_3^3$ rather than by $\hbar/(\ep_1-\ep_2)$.
 
Let us recall once again that the violation of the minimal work principle
is due to common influence of the following factors:
\begin{itemize}

\item There is a level-crossing: a more populated state goes to higher
energies, while a less populated one goes to lower energies.

\item For slow processes there are no transitions between various
energy levels: $\aad_{kn}(\tf)=\delta_{kn}$.

\item For not very slow processes there do occur transitions:
$a_{kn}(\tf)\not =\delta_{kn}$. They cost less work.

\end{itemize}

Note as well that when the external
fields $h_1(s)$ and $h_3(s)$ in Eq.~(\ref{sos}) are such that
$\sqrt{h_3^2(s)+h_1^2(s)}$ is a smooth function for all real $s$,
there are no crossings of eigenvalues and (\ref{2L}) is valid.  In the
example of level-crossing given above, one has
$\sqrt{h_3^2(s)+h_1^2(s)}\propto |s|$ for small $s$.
Note that if level-crossing is absent, the transition probability
$|a_{12}(\tf)|^2$ is small as $O(e^{-\tau})$, since the integral in 
Eq.~(\ref{w3}) is determined by the extremal point of (\ref{kra})
which is now complex; see \cite{pechu} for more details.


\subsection{Fast processes.}

One can calculate $|a_{kn}(\tf)|^2$ yet in another limiting case,
where the characteristic time $\tau$ is very short, while the change
in $h_1(t)$ and $h_3(t)$ is finite and there is the level crossing at
$t=0$.

It is well-known \cite{standardad} that in this limit energy changes
can be calculated with help of the
frozen initial state of S. For the present
situation this leads from Eq.~(\ref{sosend}) to
\BEA
&&|a_{12}(\tf)|^2=|a_{21}(\tf)|^2\nonumber\\
&&=|\langle 1,t_{\rm f}\,|\,2,t_{\rm
i}\rangle|^2 =\sin^2\half[\theta(\tf)-\theta(\ti)], 
\EEA
and thus to
\BEA
\Theta_1=(p_1-p_2)\sin^2\half[\theta(\tf)-\theta(\ti)], 
\EEA
which is
again positive. This demonstrates that even very fast processes (i.e.,
$\tau\to 0$) can be more optimal than slow ones. It is
conceivable that violations of the minimal work principle are maximal
for some finite $\tau$. This expectation is confirmed below.

\subsection{ Exactly solvable model with level crossing.}
\label{exact}

The above results obtained by perturbation theory will now be corroborated
on an exactly solvable displaying level-crossing.

Consider a two-level system with Hamiltonian 
\BEA 
\label{navu} 
H(s)=\hbar\omega 
\left(\begin{array}{rr} 
s\cos^2 s & ~~\half s\sin 2s \\ 
\\ 
\half s\sin 2s & ~~s\sin ^2s 
\end{array}\right), \qquad s=\frac{t}{\tau}, 
\EEA 
where $\tau$ is the characteristic time-scale, $s$ is the reduced
time, and where $\omega$ is a coupling constant.  The model belongs to
the pool of exactly solvable driven two-level systems.  We learned on
it from Ref.~\cite{hage}, where some asymptotic features of its
solution were studied.

The eigenvectors
\BEA
|1,s\rangle=
\left(\begin{array}{r}
\cos s\\
\\
\sin s
\end{array}\right),\qquad
|2,s\rangle=
\left(\begin{array}{r}
-\sin s\\
\\
\cos s
\end{array}\right).
\label{sosend1}
\EEA
of $H(s)$ correspond to the eigenvalues 
\BEA
\ep_1(s)=\hbar \,\omega\, s,\qquad \ep_2(s)=0,
\EEA
respectively. It is seen that the energy levels cross, when 
$s=t/\tau$ passes through zero.

Eqs.~(\ref{shaman}) read for the present case:
\BEA
\label{bund1}
&&\frac{\d a_{11}(s)}{\d s}=e^{i\omega\tau s^2/2}\,a_{12},\\
&&\frac{\d a_{12}(s)}{\d s}=-e^{-i\omega\tau s^2/2}\,a_{11}.
\label{bund2}
\EEA
These equations can be solved exactly in terms of hypergeometric functions
\BEA
a_{11}(s)= c_1 H_{\rm e}\left[-\frac{i}{\omega\tau},\, s\chi\right]+
c_2F_1\left[\frac{i}{2\omega\tau},\,\frac{1}{2}, \,s^2\chi^2\right],
\EEA
\BEA
a_{12}(s)&&= \frac{c_1}{\chi^2} ~e^{\frac{i\pi}{4}-s^2\chi^2}~
H_{\rm e}\left[-1-\frac{i}{\omega\tau},\, s\chi\right]
\nonumber\\
&&-c_2 \,s\, e^{-s^2\chi^2}
\,F_1\left[1+\frac{i}{2\omega\tau},\,\frac{3}{2}, \,s^2\chi^2\right],
\EEA
where $H_{\rm e}$ and $F_1\equiv\,_1F_1$ are, respectively, 
Hermite and hypergeometrical
functions \cite{math}, and where
\BEA
\chi\equiv e^{\frac{i\pi}{4}}\,\sqrt{\frac{\omega\tau}{2}}.
\EEA
The integration constants $c_1$ and $c_2$ are determined 
from the initial conditions
$a_{11}(s_{\rm i})=1$ and $a_{12}(s_{\rm i})=0$.

In the slow limit $\omega \tau\gg 1$, the transition probability can be
calculated via 
the first-order adiabatic perturbation theory result (\ref{w3})
\BEA
\label{bund3}
a_{12}(s_{\rm f})=-\int_{s_{\rm i}}^{s_{\rm f}}\d s\, 
e^{-i\,\omega\,\tau s^2/2}.
\EEA
With help of the saddle-point method one gets for (\ref{bund3}):
\BEA
\label{tutu}
|a_{12}(s_{\rm f})|^2=\frac{2\pi}{\omega\tau},
\EEA
in agreement with (\ref{w77})~
\footnote{To make a detailed comparison between Eqs.
(\ref{tutu}) and (\ref{w77}), substract from $H(s)$ (defined by 
(\ref{navu})) an irrelevant factor $\frac{1}{2}{\rm tr}\,H(s)$,
and note from Eqs.~(\ref{sos}, \ref{durman}) that
the following relations are valid for the present model:     
$\alpha_3=\frac{1}{2}\,\hbar\omega$, $\alpha_1=\hbar\omega$.
Substituting them into Eq.~(\ref{w77}) one gets Eq.~(\ref{tutu}).
}. The characteristic time for the slow process is seen to be
$1/\omega$.

Recalling Eq.~(\ref{bek}), one deduces from Eqs.~(\ref{suomi1}, \ref{suomi2}) 
for the present model:
\BEA
W-\Wad=\hbar\omega s_{\rm f}\,|a_{12}(s_{\rm f})|^2\, \tanh
(\half\beta\hbar\omega\, s_{\rm i}\,), 
\label{tatar}
\EEA
To have level-crossing we take $s_{\rm i}<0$, $s_{\rm f}>0$.
Eq.~(\ref{tatar}) then indeed predicts 
\BEA
W-\Wad<0.
\EEA
 
\begin{figure}[bhb] 
\vspace{-0.3cm} 
\includegraphics[width=5cm]{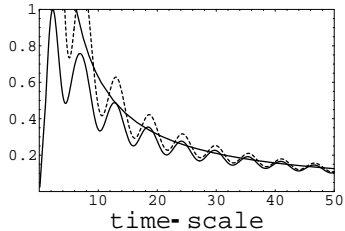} 
\vspace{-0.3cm} 
\caption{Amplitude $|a_{12}(s_{\rm f})|^2$ versus the time-scale 
$\tau$ for $s_{\rm i}=-1.5$, $s_{\rm f}=1.5$ and $\omega=1$. 
Full oscillating curve: the exact value which can reach unity.
Dotted curve: result from a first-order adiabatic perturbation 
theory. The smooth curve presents a saddle-point approximation 
(\ref{tutu}).} 
\label{bobo} 
\end{figure}

\begin{figure}[bhb]
\vspace{-0.1cm}
\includegraphics[width=5cm]{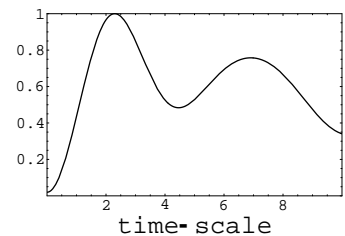}
\vspace{-0.3cm}
\caption{
The magnified version of Fig.~\ref{bobo}: The exact value of 
$|a_{12}(s_{\rm f})|^2$ versus the characteristic time-scale
$\tau$ for $s_{\rm i}=-1.5$, $s_{\rm f}=1.5$, and $\omega=1$.
}
\label{f2}
\end{figure}

It is seen that violations of the minimal work principle exist for
$s_{\rm f}>0$, and they are maximal for $|a_{12}(s_{\rm f})|^2\to 1$.
This is seen to be the case in Figs. \ref{bobo}, \ref{f2} for some
$\tau$ near $\tau=1$.  Note from Fig. \ref{bobo} that both the
first-order perturbation theory result (\ref{bund3}) and the
saddle-point approximation to it given by Eq.~(\ref{tutu}) are
adequately reproducing $|a_{12}(s_{\rm f})|^2$ for $\tau \gtrsim
10$. Moreover, the first-order perturbation theory is seen to provide
an upper bound for the exact expression and predicts the appearance of
large oscillations around $\tau\sim 1$.

For $\tau\to 0$, $|a_{12}(s_{\rm f})|^2$ goes to its value
$\sin^2(s_{\rm f}-s_{\rm i})$ predicted by the sudden perturbation
theory. This amounts to $|a_{12}(s_{\rm f})|^2=0.01991$ for the
situations in Figs.~\ref{bobo}, \ref{f2}.

Thus, all the basic conclusions drawn from the adiabatic perturbation
theory are confirmed by this exactly solvable model.

\subsection{Many-level systems.}

Let S have a finite number of levels, and assume that
two of them cross.  For
quasi-adiabatic processes ($\tau$ is large but finite) and
analytically varying Hamiltonian $H(t)$, the transition probability
between non-crossing levels is exponentially small \cite{pechu}, while,
as we saw, it has power-law smallness for two crossing levels.  Then
one neglects in (\ref{suomi0}) the factors $|a_{k\not =n}(\tf)|^2$
coming from any non-crossed levels $k$ and $n$, and the problem is
reduced to the two-level situation. Thus already one crossing suffices
to detect limits of the minimal work principle (provided, of course,
that the crossed levels are sufficiently populated initially).

The reduction to the two-level situation takes place also in a
macroscopic system which has few discrete levels located below a
continuous spectrum and separated from the latter by a finite gap.  It
is known, see e.g. in Ref.~\cite{avron}, that the transitions between
these discrete levels and the continuous part of the spectrum vanish
exponentially for quasi-adiabatic processes.  These discrete levels thus
decouple from the rest of the spectrum and the problem returns to an effectively
two-level situation.

\section{ Cyclic processes and reversibility.}  
\label{cyclic}

The above results do not imply any violation of the second law in
Thomson's formulation \cite{thomson1,thomson2,thomson3}: no work is
extracted from S during a cyclic process, $W_{\rm c}\geq 0$.  We
illustrate its general proof obtained in
Refs.~\cite{thomson1,thomson2,thomson3,ku} in the context of the level
crossing model given by Eqs.~(\ref{sos}--\ref{sosend}).

Assume that the trajectory $R(t)=(h_1(t),h_2(t)\,)$ described there is
supplemented by another trajectory $R'(t)$ which brings the parameters
back to their initial values $(h_1(\ti), h_3(\ti)\,)$ so that the
overall process $R+R'$ is cyclic. If $R'$ crosses the adiabatic energy
levels backwards, then at the final time $\tf'$ of the full cyclic
process $R+R'$ one has
\BEA
\label{domb1}
\ep_1(\tf')=\ep_1(\ti),\quad
\ep_2(\tf')=\ep_2(\ti). 
\EEA
Together with 
\BEA
|a_{12}(\tf')|^2=|a_{21}(\tf')|^2=1-
|a_{11}(\tf')|^2=1-|a_{12}(\tf')|^2,
\nonumber
\EEA
and 
Eqs.~(\ref{suomi0}, \ref{abu}) as applied to the full cyclic process
(i.e., changing in them $\tf\to\tf'$),
Eq.~(\ref{domb1}) implies:
\BEA
W_{\rm c}&&=|a_{12}(\tf')|^2(p_1-p_2)[\ep_2(\ti)-\ep_1(\ti)]\nonumber\\
&&\geq \Wad_{\rm c}=0.
\label{trombon}
\EEA

Eq.~(\ref{trombon}) confirms the intuitive expectation that
non-adiabatic process are less optimal.  In particular, this is valid
if $R'=R_{\rm mr}$ is exactly the same process $R$ moved backwards
with the same speed (mirror reflection).  Then $\Wad_{\rm c}=0$ means
that the original process
$R$ is reversible: According to the standard
thermodynamical definition \cite{landau,balian,perrot}, a process $R$
is reversible, if after supplementing it with its mirror reflection
process $R_{\rm mr}$, the work done for the total cyclic process
$R+R_{\rm mr}$ is zero.

If $R'$ does not induce another level crossing, i.e., if $h_1(s)$ and
$h_2(s)$ in Eq.~(\ref{sos}) do return to their initial values without
simultaneously crossing zero, then the levels are interchanged
\BEA
\ep_1(\tf')=\ep_2(\ti),\quad
\ep_2(\tf')=\ep_1(\ti), 
\EEA
and this time Eqs. (\ref{suomi0}, \ref{abu}) imply
\BEA
&&\Wad_{\rm c}=(p_1-p_2)[\ep_2(\ti)-\ep_1(\ti)],\\
&&\Wad_{\rm c}\geq W_{\rm c}=(\,1-|a_{12}(\tf')|^2\,)\,\,\Wad_{\rm c}>0.
\EEA
It is seen that in contrast to the situation described by
Eq.~(\ref{trombon}), non-adiabatic processes are more optimal if
$R+R'$ contains one level-crossing (or an odd number of them).  We
thus have found here a violation of the minimal work principle for a
cyclic process.

\section{Conclusion.}
\label{conco}

This paper was devoted to one of the persistent questions in
statistical physics: whether the second law is valid for finite
systems, and if it is not valid what are the possible scenarios of its
violation. The proper way to answer this question is to take a
formulation of the second law which has a clear meaning for finite
systems, and to study it from the first principles of quantum
mechanics, without invoking any postulate.

Along these lines we have studied the minimal work principle for finite
systems coupled to external sources of work.  The principle states that
the work done on an (initially) equilibrium system during a thermally
isolated process is minimal for the smallest speed of the process.  As
compared to other formulations of the second law, this principle has a
direct practical meaning as it provides a recipe for reducing energy
costs of various processes. We gave its general proof for finite
systems which starts from first principles of quantum mechanics and avoids
the usual lore associated with the second law (chaos, thermodynamic limit,
coarse-graining, various definitions of entropy).
We have also shown that it
may become limited if there are crossings of adiabatic energy levels:
optimal processes need to be neither slow nor reversible. 
Already one crossing suffices to note violations of the principle.
If this is the case, the optimal process occurs at some finite,
system-dependent speed.

Level-crossing was observed, e.g., in molecular and chemical
physics \cite{david,chem}.  It is not a rare effect \cite{rare}: if the
number of externally varied parameters is larger then two, then for
typical spectra level crossings are even much more frequent than
avoided crossings \cite{rare}.  It is possible that the presented
limits of the minimal work principle may serve as a test for level
crossings.

Together with the universal validity of Thomson's formulation of the
second law \cite{thomson1,thomson2,thomson3,ku}, the limits of the
principle imply that the very equivalence between various formulations
of the second law may be broken for a finite system coupled to
macroscopic sources of work: different formulations are based on
different physical mechanisms and have different ranges of validity.
Similar results on non-equivalence of various formulations of the
second law were found in Ref.~\cite{AN,NA}, where for a quantum
particle coupled to a macroscopic thermal bath, it was shown that some
formulations, e.g., the Clausius inequality and positivity of the energy
dispersion rate, are satisfied at sufficiently high temperatures of
the bath, but can be invalid at low temperatures, that is, in the
quantum regime. The physical mechanism responsible for this
is the formation of a cloud of bath modes around the particle, well
known in case of Kondo-cloud and polaron-cloud, but more general.

There are still many issues to consider before the minimal work
principle and its limits can be said to be properly understood. In
particular, we need a better description for the transition 
from finite to macroscopic systems. It might also be of interest to
find an explicit example of a macroscopic system displaying limits of
the principle. A separate problem is to study the minimal work
principle in the (semi)classical limit of finite quantum systems. This
problem is special due to the fact that the limit of slow processes
need not commute with the classical limit \cite{berry}.

We, however, had first to understand that limits of minimal work principle
do exist in principle and sometimes in practice, and this is the main
message of the present paper.

\section*{Acknowledgments.}

We thank R. Balian for his constructive criticisms
and K. Sekimoto for comments.


The work of A.E. A is part of the research programme of the Stichting voor 
Fundamenteel Onderzoek der Materie (FOM, financially supported by 
the Nederlandse Organisatie voor Wetenschappelijk Onderzoek (NWO)).

\appendix

\section{Limits to the second law.}
\label{disco}

As was stressed in the Introduction, one of basic purposes of this
paper was to understand limits of the minimal work principle, which is
a particular formulation of the second law.  We therefore consider it
necessary to clarify what in general we mean by ``limits to the second
law''.

It is first of all important that one is dealing with a concrete
formulation of the second law, which, within the studied situation,
has a clear conceptual and operational meaning. For example, Thomson's
formulation of the second law as applied to finite systems is certainly of
that type, and so is the minimal work principle.  Both these
formulations of the second law operate with the concept of work, which
| in contrast to entropy | is defined unambiguously for finite
systems at all times.

If within the studied situation we find the statement not satisfied |
e.g., the minimal work principle need not be satisfied in the presence
of level crossings | then we encountered a limit of that particular
formulation.  Note that for finite systems it will not be legitimate
to speak about limits of any entropic statement of the second law, since
entropy for a finite system does not have a clear and well-accepted
physical meaning. One can define here various entropies (especially in
the quantum situation) and it is not clear with respect to which one
the corresponding statements of second law are to be formulated. Here
one encounters limits of the concept of entropy rather than limits of
the second law. It is only in the macroscopic situation that
the coarse-grained entropy becomes a meaningful quantity
on sufficiently long time-scales.

In the context of limits of the second law there is an opinion that
fluctuations of various quantities | e.g., fluctuations of work,
provided they can be sensibly defined | provide violations of the
second law. This opinion is supported by an observation that for a
typical system in the thermodynamical limit the fluctuations vanish
and, for example, Thomson's formulation in this limit is a statement
on energy difference of a macroscopic system (i.e., in the definition
(\ref{work}) of work one need not take averages over different systems
or different realizations).  According to this opinion, the second law
is formulated as a statement on random quantities (e.g., on the random
quantity work), and once they can fluctuate for finite systems, this
gives violations of the second law.

It is true that when the second law was first deduced in the XIX
century, it was formulated for a single closed system, in a way
resembling the laws of ordinary mechanics. However, already in the
beginning of XX century it was clearly understood
\cite{gi,epstein,tolman,kemble} that this law has only a statistical
character and refers to averages over an ensemble of identically
prepared systems, rather than to a single system.  This viewpoint became
widely accepted when the first robust observations of fluctuations were
made  \cite{epstein}. Together with theoretical works of Boltzmann in
kinetic theory of gases and of Smoluchowski, Fokker, Planck and
Einstein in physics of Brownian motion, they formed a consistent picture
of the second law as emerging from micro-physics through averaging over
fluctuations.  A detailed summary of this activity is presented in the 1937
book by Epstein \cite{epstein}, while Tolman in 1938 \cite{tolman} and
Kemble in 1939 \cite{kemble} discuss theoretical aspects of the situation.
Since then, this understanding of the second law entered into several
modern books of statistical physics and
thermodynamics \cite{landau,balian}.

It is also true that when the statistical character of the second law
was not yet widely accepted, several known scientists made statements
on violations of the second law by fluctuations \cite{jo}.  One of
these citations is by Maxwell `` the second law is drawn from our
experience of bodies consisting of immense number of molecules.... it
is continually being violated in any sufficiently small group of
molecules. As the number ... is increased .. the probability of a
measurable variation may be regarded as practically an
impossibility. ''  If one cites this quotation, one should keep in
mind that Thomson's formulation of the second law is perfectly valid
for any ``small group of molecules''. As we saw in the present paper,
the same | modulo level-crossing | concerns the minimal work
principle.

In summary, fluctuations do not provide violations of the second law, since
this law is formulated with respect to averages. For a recent discussion on
this point see \cite{ANcontra}.

\end{document}